\documentclass[twocolumn,trackchanges]{aastex631}
\usepackage{graphicx}	% Including figure files
\usepackage{amsmath}	% Advanced maths commands
\usepackage{hyperref}
\usepackage{xcolor}
\begin{document}

\title{Search for young stellar objects within 4XMM-DR13 using CatBoost and SPE}

\author{Xiangyao Ma}
\affiliation{College of Physics, Hebei Normal University, Shijiazhuang 050024, China}
\affiliation{National Astronomical Observatories, Chinese Academy of Sciences, Beijing, 100101, China}

\author[0000-0002-6610-5265]{Yanxia Zhang}
\affiliation{National Astronomical Observatories, Chinese Academy of Sciences, Beijing, 100101, China}

\author{Jingyi Zhang}
\affiliation{National Astronomical Observatories, Chinese Academy of Sciences, Beijing, 100101, China}

\author{Changhua Li}
\affiliation{National Astronomical Observatories, Chinese Academy of Sciences, Beijing, 100101, China}

\author{Zihan Kang}
\affiliation{National Astronomical Observatories, Chinese Academy of Sciences, Beijing, 100101, China}

\author{Ji Li}
\affiliation{College of Physics, Hebei Normal University, Shijiazhuang 050024, China}

\correspondingauthor{Yanxia Zhang, Jingyi Zhang, Changhua Li}
\email{zyx@bao.ac.cn, jyzhang@bao.ac.cn, lich@bao.ac.cn}

\begin{abstract}

Classifying and summarizing large data sets from different sky survey projects is essential for various subsequent scientific research. By combining data from 4XMM-DR13, SDSS DR18, and CatWISE, we formed an XMM-WISE-SDSS sample that included information in the X-ray, optical, and infrared bands. By cross-matching this sample with datasets from known spectral classifications from SDSS and LAMOST, we obtained a training dataset containing stars, galaxies, quasars, and Young Stellar Objects (YSOs). Two machine learning methods, CatBoost and Self-Paced Ensemble (SPE), were used to train and construct machine learning models through training sets to classify the XMM-WISE-SDSS sample. Notably, the SPE classifier showed excellent performance in YSO classification, identifying 1102 YSO candidates from 160,545 sources, including 258 known YSOs. Then we further verify whether these candidates are YSOs by the spectra in LAMOST and the identification in the SIMBAD and VizieR dabtabases. Finally there are 412 unidentified YSO candidates. The discovery of these new YSOs is an important addition to existing YSO samples and will deepen our understanding of star formation and evolution. Moreover we provided a classification catalog for the whole XMM-WISE-SDSS sample. 

\end{abstract}



\section{Introduction} \label{sec:intro}
The young stellar object (YSO) dataset serves as a crucial resource for elucidating the star formation and evolutionary history of galaxies, offering insights into the diverse processes involved in star formation. Given that we are unable to observe the formation of a star over extended periods, we can only infer the formation processes of various stars by analyzing YSO data at different evolutionary stages \citep{chiu2021searching}. The evolutionary stages of stars were first studied and classified by \citet{adams1987spectral} and \citet{Lada1987}, and then further improved by \citet{andre1993submillimeter} and \citet{greene1994further}. \citet{allen2004infrared} mentioned that observing the spectral energy distribution (SED) in the infrared spectrum can identify the evolution stage from the pre-satellite core stage to the main sequence. Based on the characteristics of their infrared spectral energy distribution, YSO can be categorized  into four classes, from Class 0 to Class III \citep{Lada1987}. The most accurate description of YSO would be based on different stages of evolution rather than categories \citep{dunham2014evolution}. According to the evolution stages of stars, it is helpful to classify YSOs.
Furthermore, young stellar objects (YSOs) exhibit significant magnetic activity, resulting in the emission of intense X-ray radiation that can effectively penetrate the surrounding dusty environments \citep{kuhn2015spatial}. High-resolution X-ray telescopes are pivotal for detecting YSOs, thereby enhancing our ability to identify them. Utilizing data from the Chandra X-ray Observatory and Gaia Data Release 2, \citet{wang2020stellar} compiled a catalog detailing the X-ray activity of approximately 6,000 stars, identifying 1,196 YSOs within this dataset.

The rapid implementation of survey projects has facilitated the collection of extensive data, enabling in-depth and comprehensive analyses. However, managing and processing these large datasets has become increasingly challenging. In recent years, the swift development of machine learning techniques, various computational algorithms, and advanced processing units has emerged as a promising solution for handling extensive data, marking an inevitable trend in the field of data analysis \citep{chiu2021searching}. Compared to traditional methodologies, classification models derived from big data training offer a convenient and effective means of identifying YSOs. Numerous studies have successfully employed these techniques to discover new YSOs. For instance, \citet{marton2016all} employed Support Vector Machines (SVM) for the recognition of Young Stellar Objects (YSOs), while \cite{marton2019identification} utilized Random Forest (RF) for the same purpose. Additionally, \citet{kuhn2021spicy-2} applied machine learning techniques to identify YSOs across various large datasets. More recently, \citet{rimoldini2023gaia} leveraged data from Gaia Data Release 3 (DR3) to propose a comprehensive classification system for all-day variable sources, which includes the classification of YSOs. The increasing application of various machine learning methods underscores their growing importance in the identification and analysis of YSOs within large datasets.

%For instance, \citet{ksoll2021measuring} utilized Support Vector Machines (SVM) and Random Forest (RF) algorithms, while \citet{cornu2020modeling} implemented Convolutional Neural Networks (CNN). Additionally, \citet{cornu2021neural} leveraged Artificial Neural Networks (ANN). 

We performed a comprehensive analysis by cross-matching the 4XMM-DR13 catalog with photometric data from SDSS DR18. Following this, we incorporated CatWISE data to create an extensive database that encompasses XMM-Newton X-ray, SDSS optical, and CatWISE infrared datasets. Utilizing the spectral databases from SDSS DR18 and LAMOST DR9, we classified the spectral types of known samples. With this integrated database and the established spectral classes as our foundation, we employed two machine learning methodologies: CatBoost and Self-Paced Ensemble (SPE) for classification. These techniques capitalize on the synergy of X-ray, optical, and infrared data. Subsequently, we conducted a comparative analysis of the classification outcomes derived from these two machine learning approaches, assessing their effectiveness in identifying and classifying astronomical objects.

The structure of this paper is organized as follows: Section 2 outlines the survey data essential for the study and the criteria for sample selection. Additionally, we present a two-dimensional feature map that highlights the importance of employing machine learning techniques for classification. Section 3 introduces the two machine learning methodologies, CatBoost and Self-Paced Ensemble (SPE), detailing their application to the samples and the corresponding evaluation metrics used for assessment. In Section 4, we present the results obtained from applying both machine learning methods to the samples, followed by a comparative discussion and further validation of the identified YSO candidates. Finally, Section 5 provides a summary of the research findings and outlines potential directions for future work.

\section{Data}
\subsection{XMM-Newton}
The XMM-Newton Scientific Operation Center (SOC) oversees the operations of the XMM-Newton satellite, the largest X-ray astronomical observation satellite launched by the European Space Agency (ESA) to date. Since its deployment on December 10, 1999, XMM-Newton has been at the forefront of high-energy astrophysics \citep{webb2020xmm}. It is renowned for its high sensitivity across a broad energy spectrum, ranging from 0.1 to 12 keV \citep{peca2024stripe}, enabling the detection of faint X-ray sources from soft to hard X-rays. Equipped with three primary scientific instruments: the European Photon Imaging Camera (EPIC), the Reflection Grating Spectrometer (RGS), and the Optical Monitor (OM) \citep{mason20244xmm}, XMM-Newton excels in delivering detailed X-ray imaging and spectral observations. These capabilities have significantly contributed to diverse research areas, including the study of black holes, neutron stars, supernova remnants, galaxy clusters, and active galactic nuclei (AGN), enriching our understanding of high-energy processes in the universe \citep{huang2023xmm,defrancesco2023nustar,eagle2023joint}. 

The 4XMM-DR13 catalog, the most comprehensive to date, includes source detections from 13,243 EPIC observations, reflecting significant enhancements over previous catalogs and spanning two decades of XMM-Newton’s contributions to X-ray astronomy. For our analysis, we primarily utilize the total band flux (f8), which encompasses the energy range of 0.2 keV to 12 keV, along with the f9 X-ray band flux, which spans from 0.5 keV to 4.5 keV, as detailed in Table~\ref{Table 2}.

\subsection{SDSS}
The 18th data release (DR18) of the Sloan Digital Sky Survey (SDSS) marks the debut of the fifth-generation SDSS-V survey. The SDSS-V consists of three survey components or ``plotters" : the Galaxy Plotter (MWM), the Black Hole Plotter (BHM), and the Local Volume Plotter \citep{sanchez2020multi}. Since its inception in 1998, SDSS has observed stars, galaxies, quasars, and various celestial bodies from the solar system to the early universe almost continuously. The primary goal of SDSS-V is to delve into the historical narrative of the Milky Way, targeting more than 6 million objects, tracing their chemical composition and revealing the inner workings of the origins of stars and planets. SDSS-V mapping interstellar gas spectra in galaxies 1000 times larger than existing samples will make a significant contribution to understanding the self-regulation mechanisms of galactic systems \citep{kollmeier2019sdss}. The SDSS-V data version contains extensive target information for two multi-target spectroscopy programs (MWM and BHM), including input catalogs and selection capabilities for many scientific targets. The MWM YSO project focuses on the pre-main sequence phase of low-mass stars, enabling exploration of key aspects of the evolution of early low-mass stars, such as where stars are born and the diffusion processes that lead to their transformation into wild stars. The MWM investigation in SDSS-V utilizes BOSS low-resolution (R$ \sim $2000) optics and APOGEE (R$ \sim $22,500) near-infrared spectroscopy \citep{almeida2023eighteenth}. The survey will produce a comprehensive dataset of more than 5 million objects in the sky, helping to map dense continuous stars in the lower dimensions of the Milky Way. High signal-to-noise ratio (S/N), medium-resolution near-infrared spectroscopy plays a crucial role in inferring stellar parameters and chemical compositions, enhancing our understanding of the main formation mechanisms of galaxies. We mainly use the information of $u, g, r, i, z$ bands of SDSS. The approximate limiting magnitudes for each band are 22.0 ($u$), 23.0 ($g$), 22.5 ($r$), 22.0 ($i$) and 20.5 ($z$).

\subsection{CatWISE}
The CatWISE2020 directory comprises a staggering 1,890,715,640 sources. These sources were meticulously curated from Wide-Field Infrared Survey Explorer (WISE) and NEOWISE measurements at 3.4$\mu$m and 4.6$\mu$m ($W1$ and $W2$), spanning from January 7, 2010 to December 13, 2018 \citep{marocco2021catwise2020}. Compared to the CatWISE preliminary catalog \citep{eisenhardt2020catwise}, CatWISE2020 is two years older, resulting in more than double the number of stars in the catalog. Notably, the total CatWISE2020 exposure surpasses that of the AllWISE catalog by six times, with a time baseline extending to 16 times that of the AllWISE catalog. Furthermore, CatWISE2020 demonstrates enhanced astrometric performance for faint stars, highlighting its superior capabilities.

\subsection{LAMOST}
LAMOST stands as the world's largest optical telescope, featuring a large aperture and a wide field of view achieved through thin mirror active optics and splicing mirror active optics technology. It boasts the capability to observe objects as faint as 20.5 magnitude within a mere 1.5 hours of exposure. Notably, LAMOST also holds the distinction of being the world's highest spectral acquisition efficiency astronomical telescope, capable of simultaneously collecting 4000 spectra. This exceptional efficiency is attributed to LAMOST's adoption of a parallel controllable dual-rotation fiber positioning scheme, utilizing 4000 fibers distributed on the focal surface with 1.75 meter diameter \citep{Luo2012,Luo2015}. LAMOST has released data for the ninth time, comprising 10,809,336 low-resolution spectra (LRS) and 8,640,738 medium-resolution spectra (MRS). These LRS are categorized into different groups, including 10,495,781 stars, 238,558 galaxies, and 74,997 quasars. 

%In this paper, we only construct known samples according to the classification of LAMOST.

\subsection{Known Samples}
Since its inception in 2009, the Million Quasars (Milliquas) catalog has stood out as the most comprehensive catalog of quasars, encompassing all the latest quasar discoveries. As of August 2, 2023, Milliquas Catalog, version 8 includes a total of 907,144 Class I QSOs and AGNs \citep{flesch2015half}. Additionally, it lists 66,026 QSO candidates identified with a 99\% probability of being quasars through combined radio/X-ray calculations. We selected type I and II quasars from Milliquas and extracted a total of 727,653 quasars. SDSS has a rich history of compiling and releasing quasar catalogues, primarily for cosmological and quasar physics research. The SDSS DR16 Quasar catalogue (DR16Q) includes 750,414 quasars \citep{lyke2020sloan}.

To compile a dataset encompassing X-ray, optical, and infrared band features, we cross-matched the 4XMM-DR13 catalog with the SDSS and CatWISE2020 databases. As demonstrated by \citet{zhang2021classification}, the integrity and contamination rates for XMM matching with SDSS at angular separations of 3, 4, 5, and 6 arc second are as follows: 71.68\% integrity versus 9.75\% contamination, 79.49\% integrity versus 16.53\% contamination, 85.55\% integrity versus 24.50\% contamination, and 90.78\% integrity versus 33.36\% contamination, respectively. For the XMM matched with WISE at separations of 3, 6, 7, and 8 arc second, the integrity and contamination rates are 56.90\% versus 0.89\%, 77.66\% versus 1.75\%, 83.39\% versus 2.01\%, and 89.14\% versus 2.08\%, respectively. The cross-matching of SDSS and 4XMM-DR9 sources achieves over 90\% integrity at 6 arc second, and approximately 90\% integrity when matching with WISE at 8 arc second. To optimize for higher integrity and lower contamination rates, we have set the cross-matching radius between 4XMM-DR13 and SDSS sources to 6 arc second, and between 4XMM-DR13 and CatWISE2020 sources to 8 arc second. We derived the XMM-WISE sample and XMM-SDSS sample, subsequently combining them based on matching IDs to obtain the XMM-WISE-SDSS sample. Known samples were cross-matched with the XMM-WISE-SDSS samples using a cross-matching radius of 3 arcsecs. Spectral identification for these samples was obtained from SDSS DR18 and LAMOST DR9. For data quality maintenance, $zWarning = 0$ was set in the DR18 SpecObjAll database during data retrieval, which indicates that no warnings or issues were flagged during the redshift measurement process. The conversion to AB magnitude for $W1$ and $W2$ involved adjustments such that $W1_{\rm AB} = W1 + 2.699$
and $W2_{\rm AB} = W2 + 3.339$, similarly the transformation to AB magnitudes for $u$ and $z$ was conducted using $u_{\rm AB} = u - 0.04$ mag and $z_{\rm AB} = z + 0.02$ mag. Additionally, the extinction correction was applied to all photometric magnitudes similar to the work of \cite{Schindler2017}. The labels of known QSO samples were obtained from the Milliquas and SDSS DR16Q catalogues, those of known stars and galaxies are from SDSS and LAMOST, and the labels of known YSOs are from the NEOWISE catalogues \citep{park2021quantifying}.

Through the above process, the known sample comprises 4459 stars, 8859 galaxies, 24,024 quasars, and 350 YSOs. Notably, the star sample excludes YSOs. We then limit $u, g, r, i$ and $z$ of known samples to 0$\sim$23 mag. Finally, the training sample contains 4111 stars, 5605 galaxies, 21,001 quasars and 258 YSOs, as shown in Table \ref{Table 1}, encompassing data from X-ray, optical, and infrared wavelengths. Table \ref{Table 1} reveals a significant imbalance in the distribution of labeled datasets, prompting us to construct a classifier fit for imbalance learning. 

\begin{table}
    \centering
    \caption{The number of the known samples.}
    \label{Table 1}
{\begin{tabular}{ll}
         \hline
         Class&  No.\\
         \hline
       Galaxy&  5605\\
         Star&  4111\\
          QSO&  21,001\\
          YSO&  258\\
         \hline
    \end{tabular}}
\end{table}

\begin{figure*}
    \centering
    \includegraphics[width=0.32\textwidth]{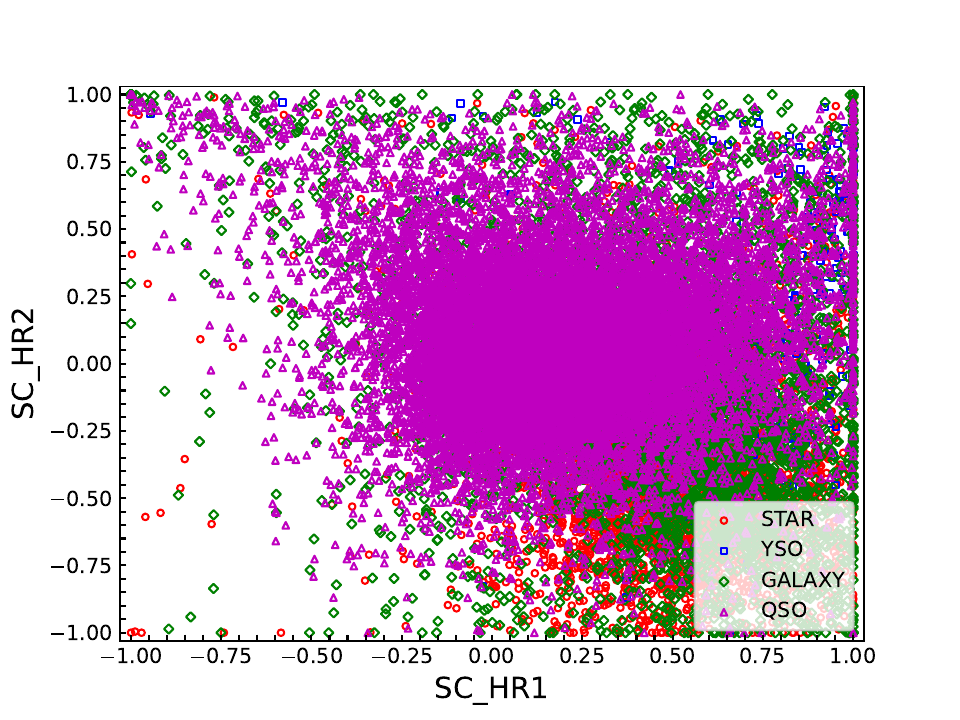}
    \hfill
    \includegraphics[width=0.32\textwidth]{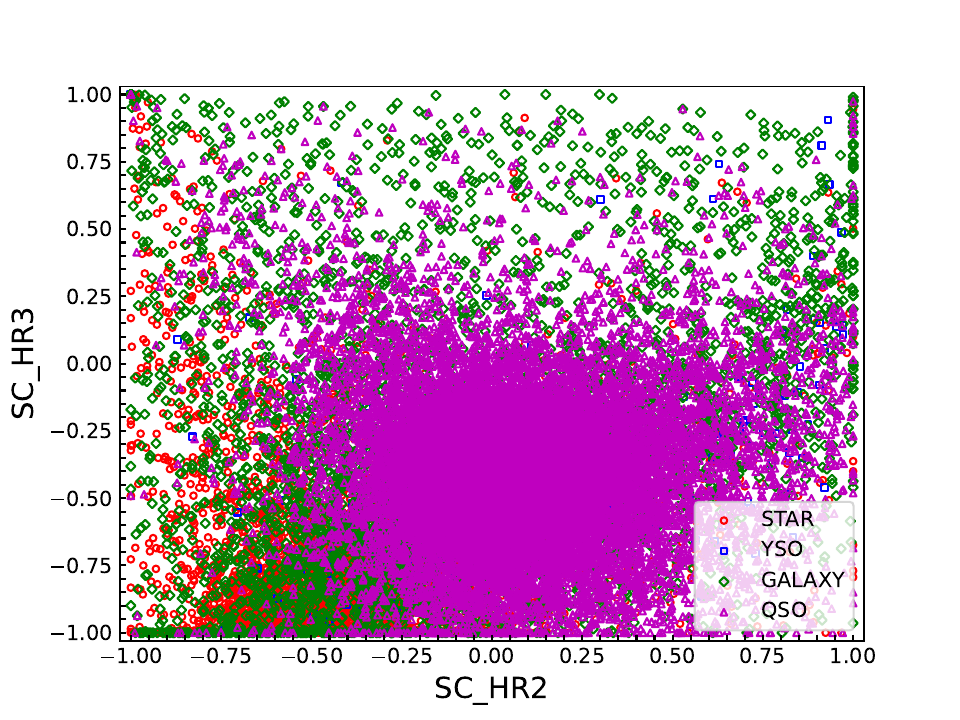}
    \hfill
    \includegraphics[width=0.32\textwidth]{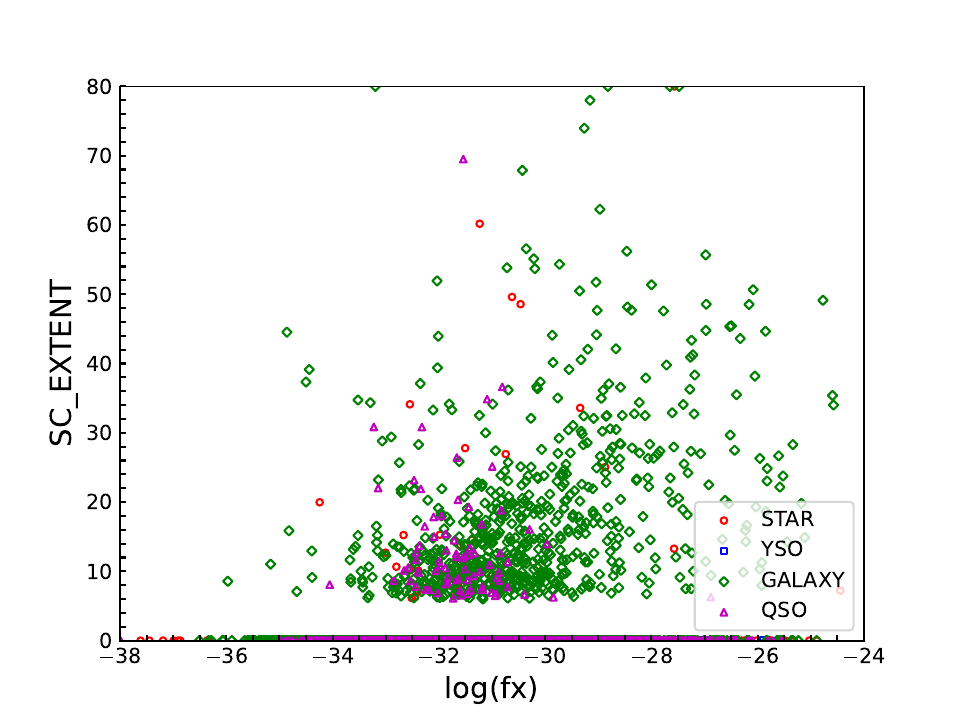}

    \vspace{0.1cm}

    \includegraphics[width=0.32\textwidth]{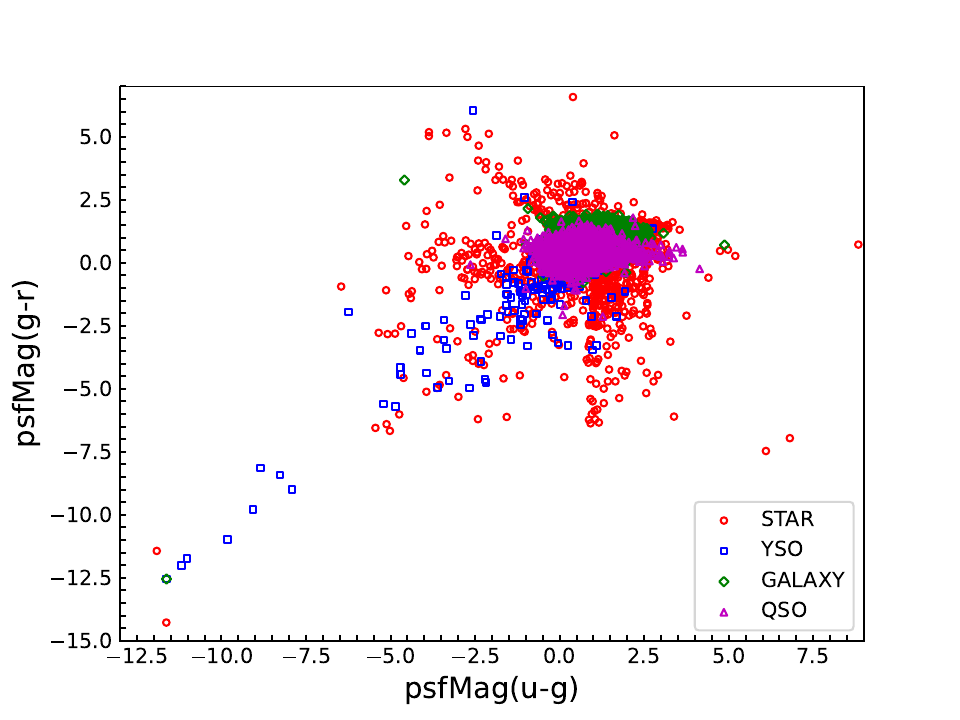}
    \hfill
    \includegraphics[width=0.32\textwidth]{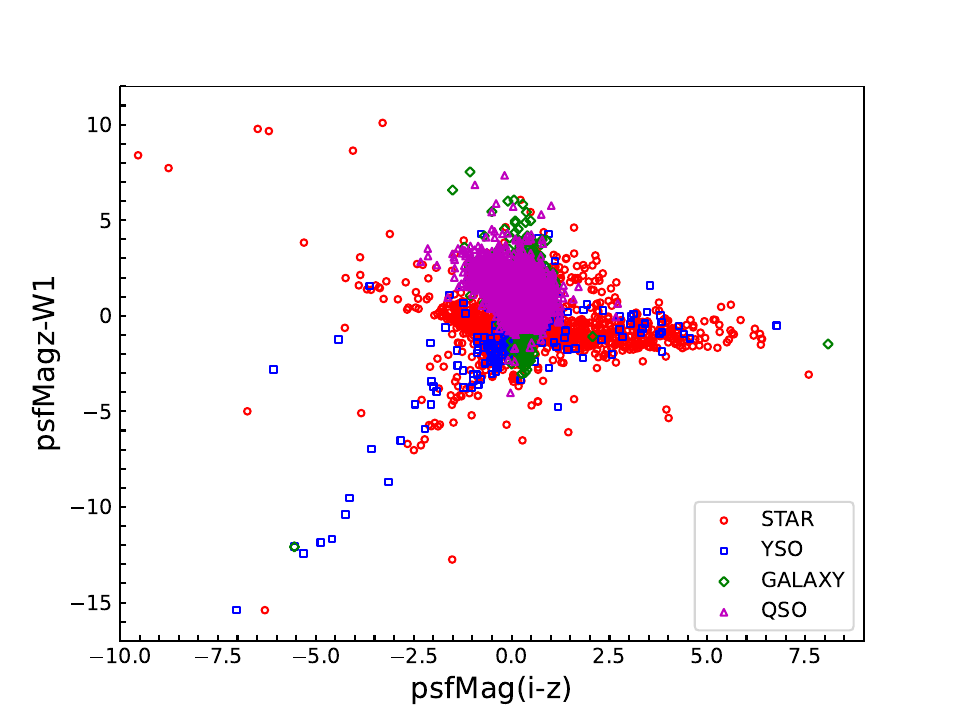}
    \hfill
    \includegraphics[width=0.32\textwidth]{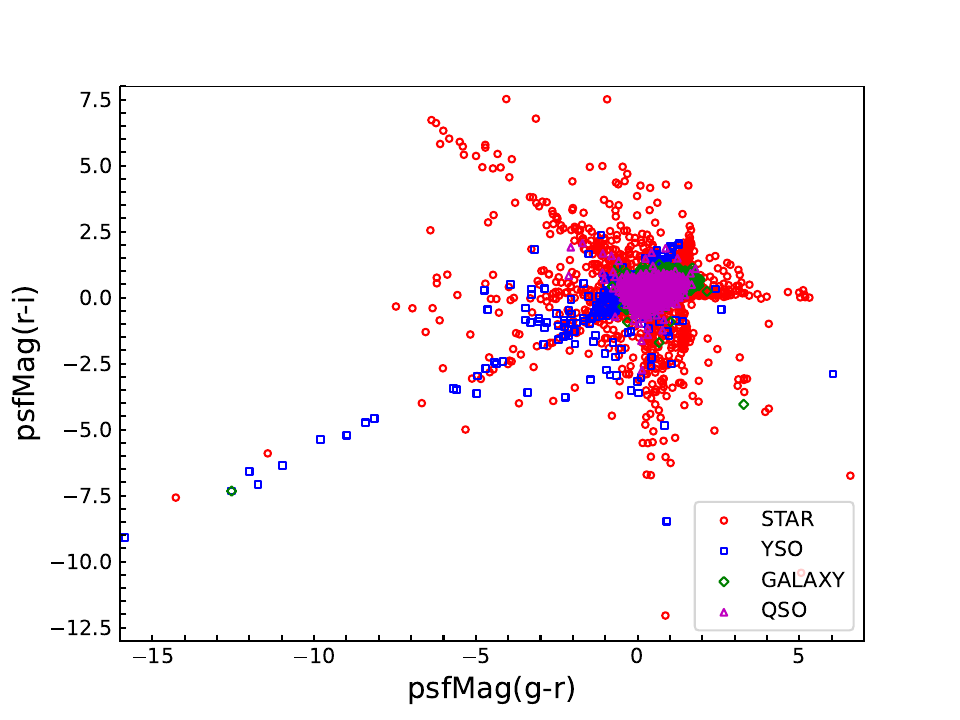}

    \vspace{0.1cm}

    \includegraphics[width=0.32\textwidth]{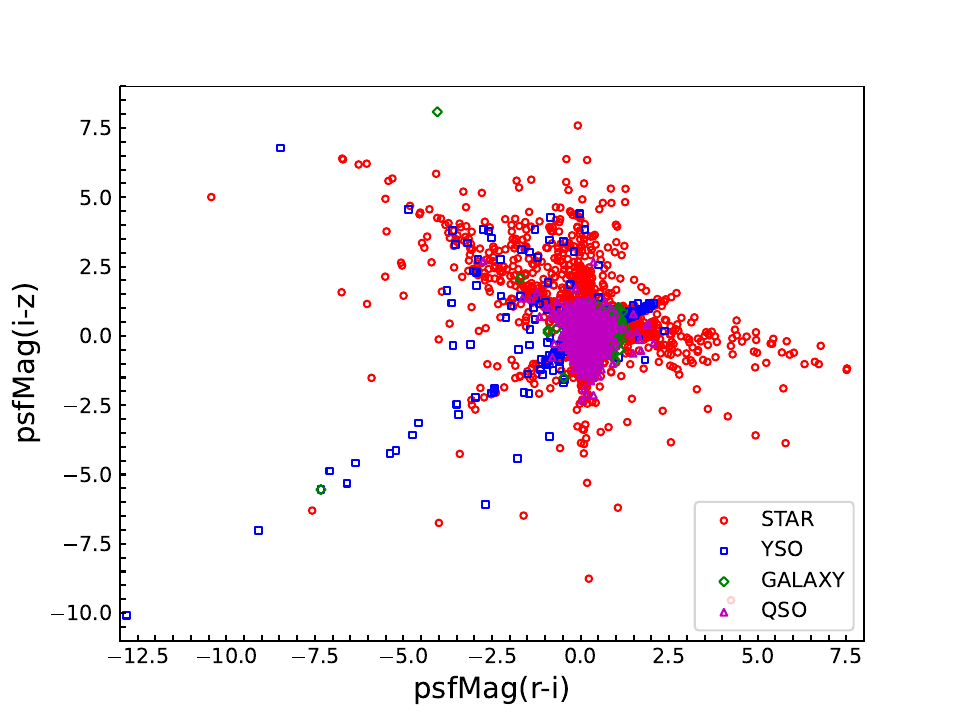}
    \hfill
    \includegraphics[width=0.32\textwidth]{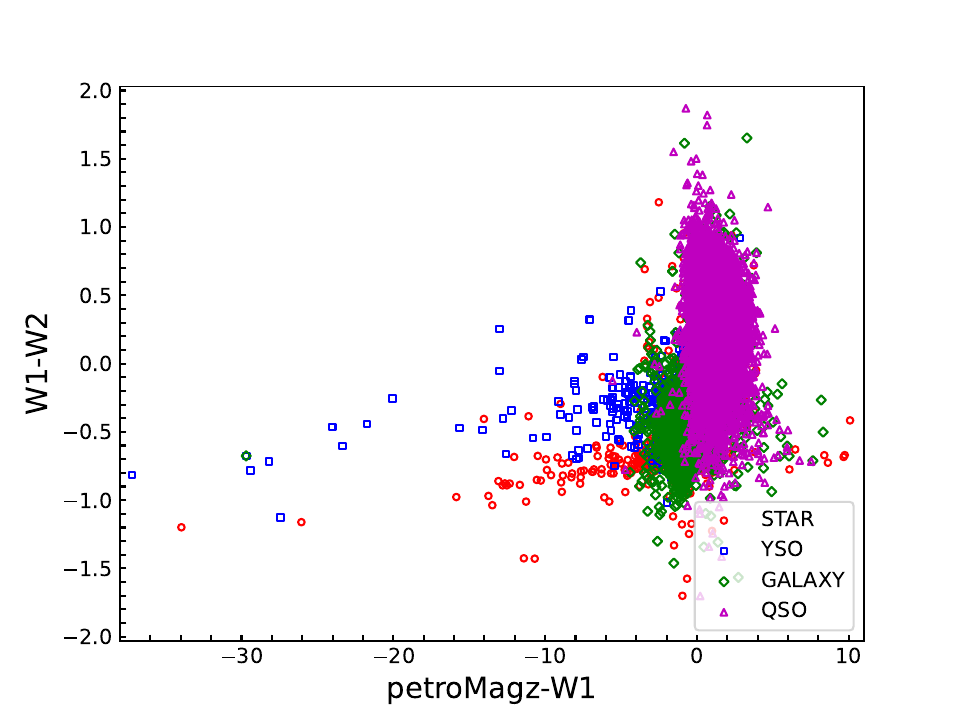}
    \hfill
    \includegraphics[width=0.32\textwidth]{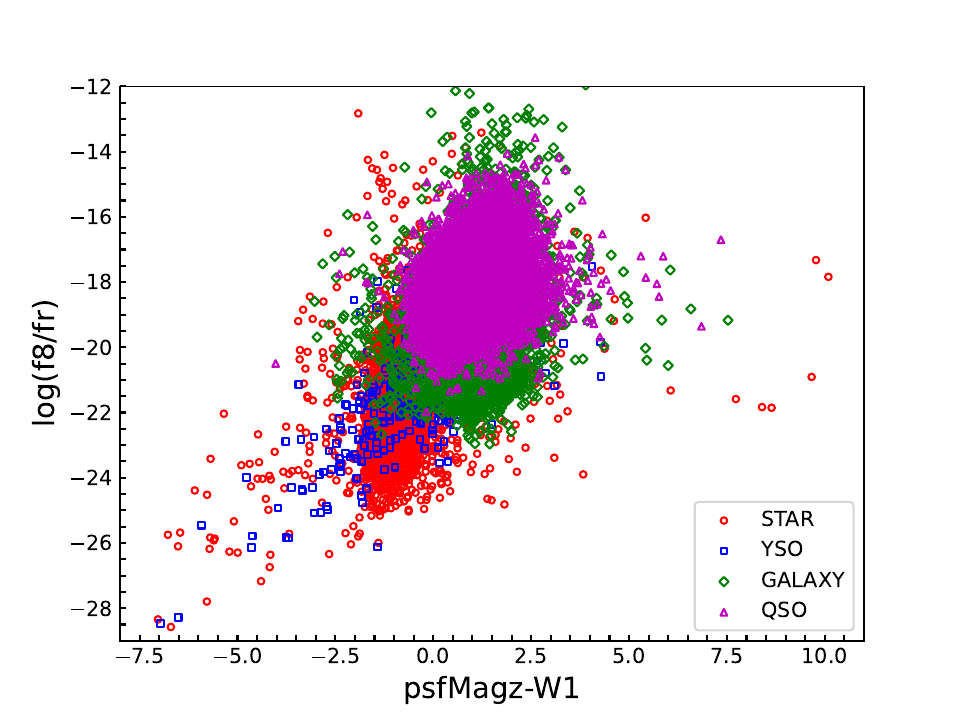}
    
    \vspace{0.1cm}

    \includegraphics[width=0.32\textwidth]{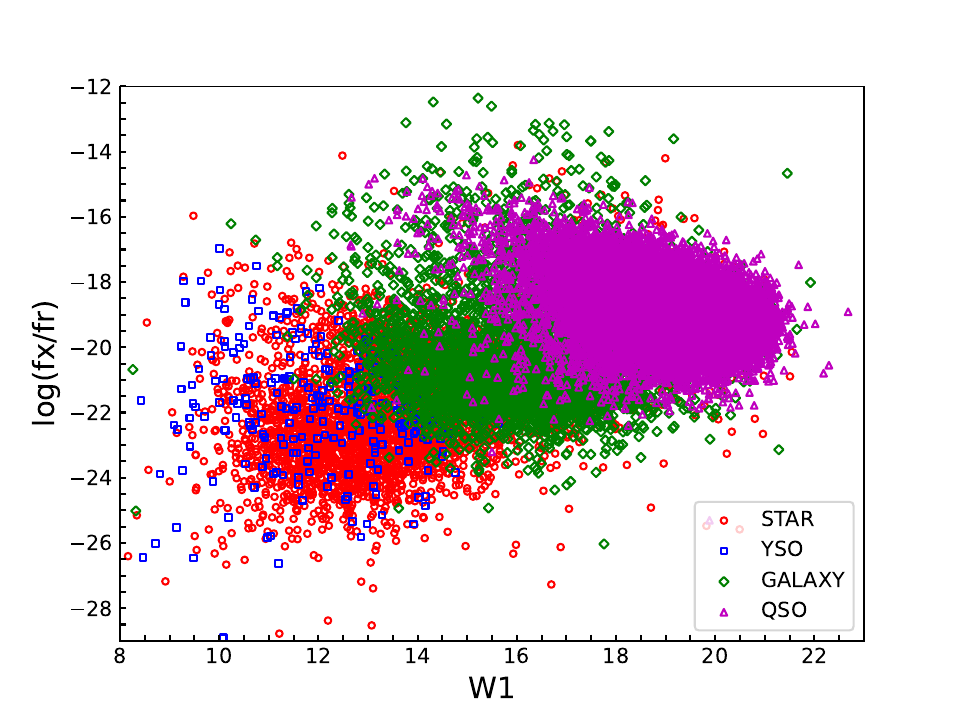}
    \hfill
    \includegraphics[width=0.32\textwidth]{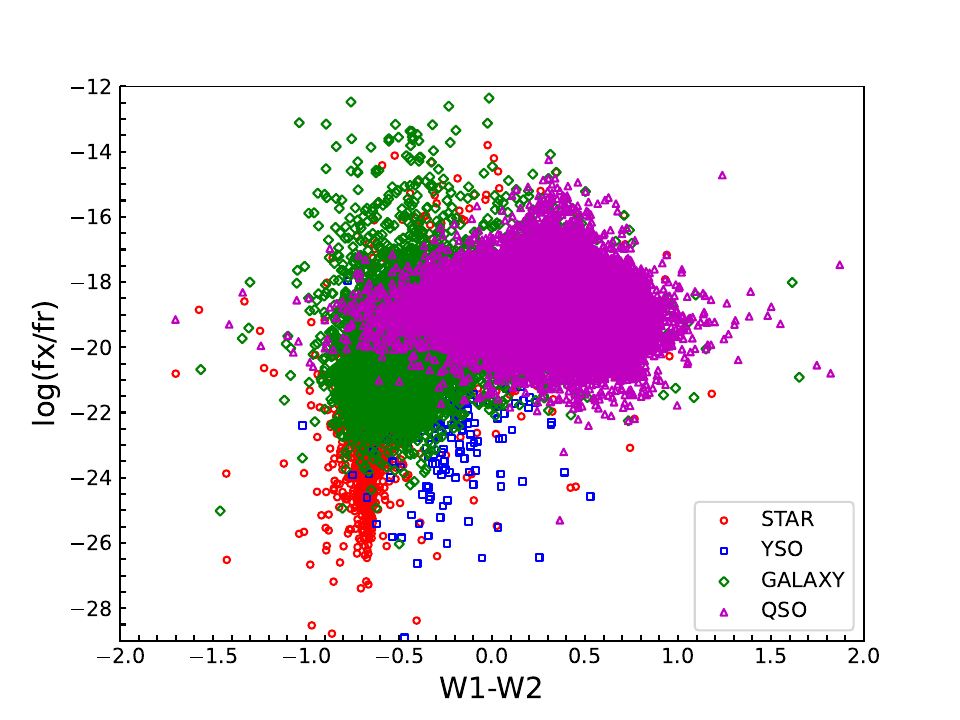}
    \hfill
    \includegraphics[width=0.32\textwidth]{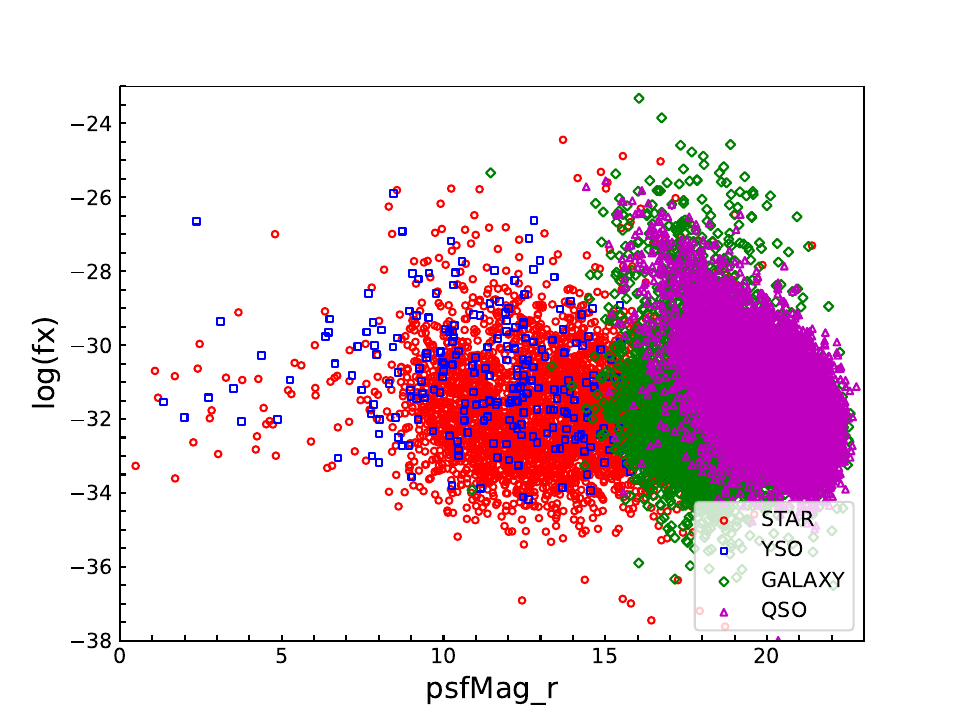}
    
    \vspace{0.1cm}

    \includegraphics[width=0.32\textwidth]{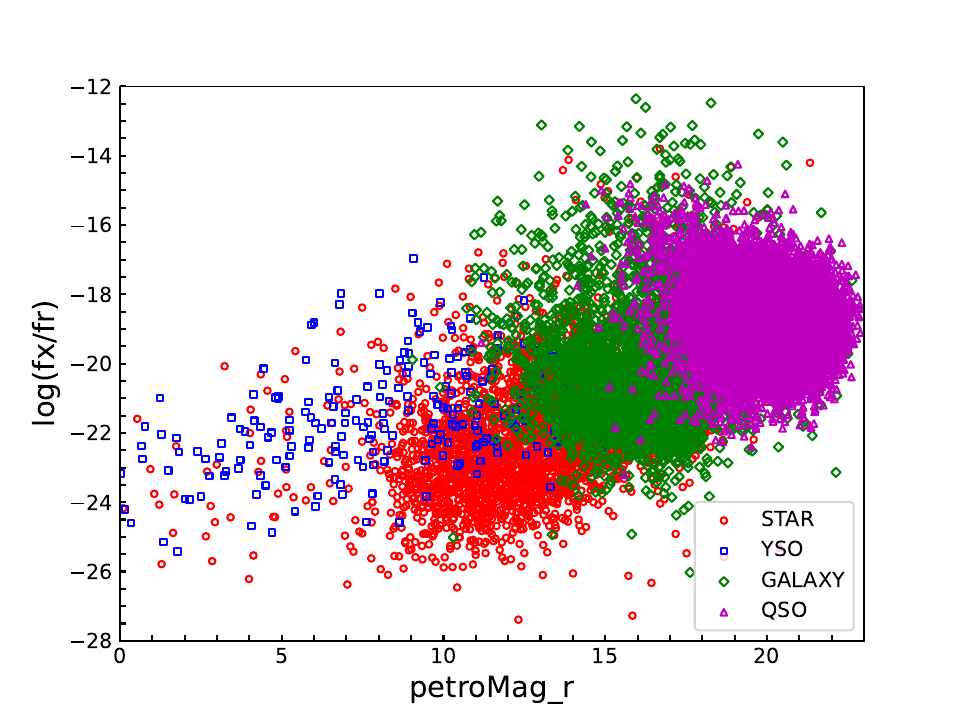}
    \hfill
    \includegraphics[width=0.32\textwidth]{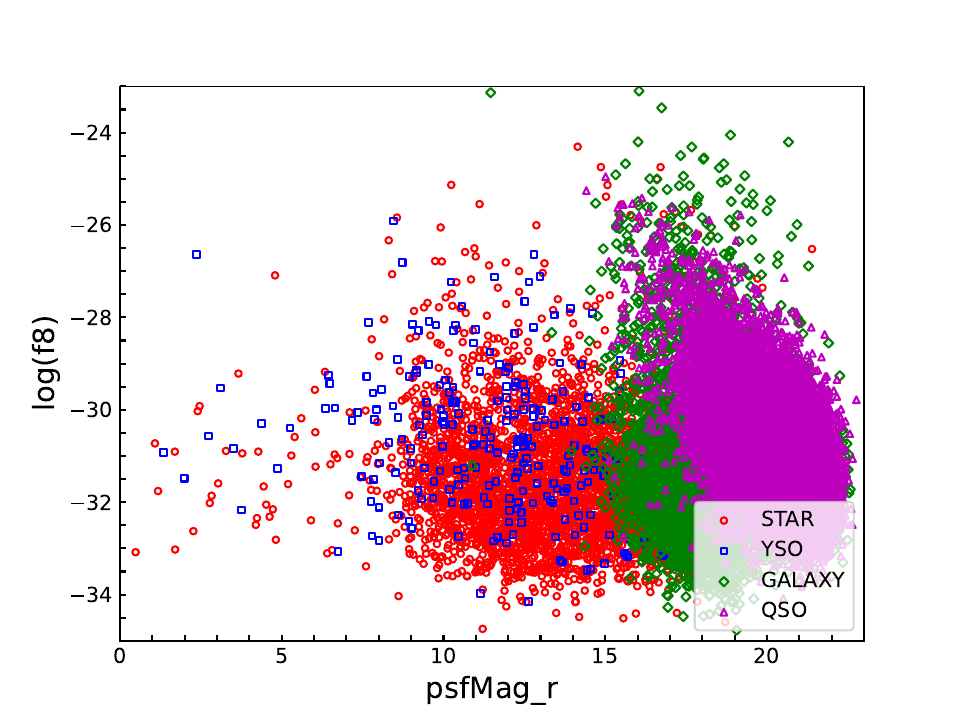}
    \hfill
    \includegraphics[width=0.32\textwidth]{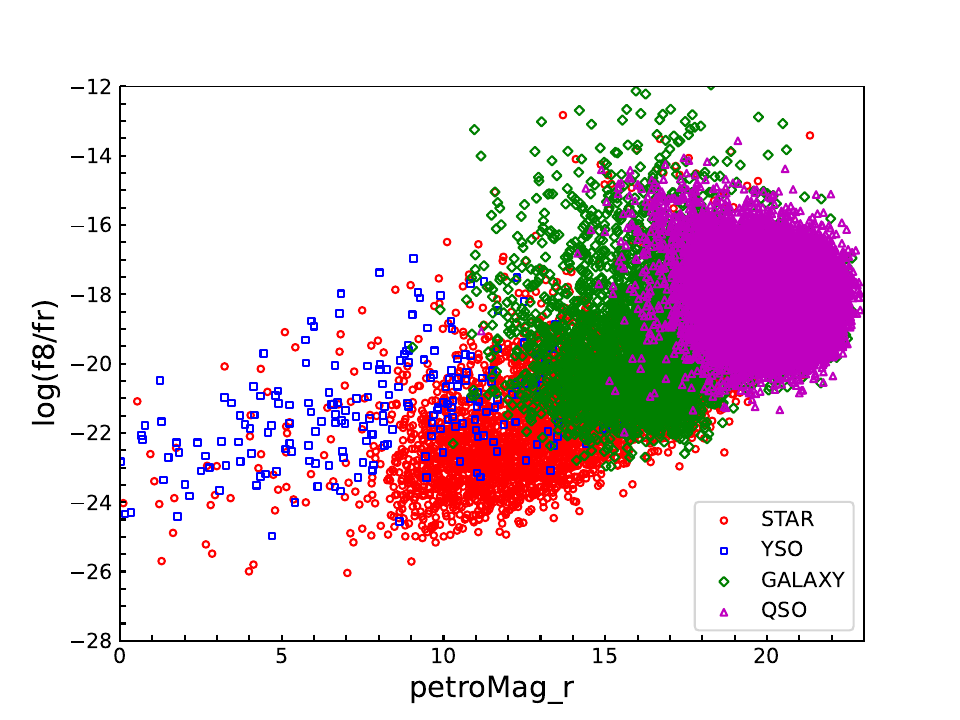}

    \caption{The distribution of stars, galaxies, quasars and YSOs in 2D spaces; the red open circles represent stars, the blue open squares for YSOs, the green open diamonds for galaxies, and the purple open triangles for quasars.}
    \label{Figure 1}
\end{figure*}

The characteristics of the sample selected for our study are outlined in Table \ref{Table 2}. Based on these data features, we generated a two-dimensional graph illustrating the relationships between these features, depicted in Figure~\ref{Figure 1}. Figure~\ref{Figure 1} indicates that distinguishing between stars, quasars, galaxies, and YSOs solely based on two attributes is not feasible. However, an observation from the figure reveals that the values of $psfMag\_r$, $W1$, and $log(f8/fr)$ for quasars and galaxies are notably larger than those for stars and YSOs; most values of $pdfMag(i-z)$, $psfMagz-W1$ and $psfMag(g-r)$ are larger for YSOs than those for quasars and galaxies. But there is not a clear separation among stars, quasars, galaxies and YSOs in any 2D diagram. This observation suggests that it may not be possible to discriminate these celestial objects only using two features, while these features do influence classification to varying degrees. Therefore, we developed optimized classification models leveraging combinations of these features to effectively classify these celestial objects by means of machine learning methods.

\begin{table*}
    \centering
    \caption{The parameters, definition, catalogues and wavebands.}
    \label{Table 2}
    \resizebox{\linewidth}{!}{%
    \begin{tabular}{llll}
         \hline
         Parameter & Definition & Catalogue & Waveband \\
         \hline
         SRCID & Source ID & XMM & X-ray band \\
         SC\_RA & Right ascension in decimal degrees & XMM & X-ray band \\
         SC\_DEC & Declination in decimal degrees & XMM & X-ray band \\
         SC\_HR1 & Hardness ratio 1 XMM X-ray band & XMM & X-ray band \\
         & Definition: hr1 = (B$-$A)/(B $+$ A), where &  & \\
         & A = count rate in energy band 0.2$-$0.5 keV &  & \\
         & B = count rate in energy band 0.5$-$1.0 keV &  & \\
         SC\_HR2 & Hardness ratio 2 & XMM & X-ray band \\
         & Definition: hr2 = (C $-$ B)/(C $+$ B), where &  & \\
         & B = count rate in energy band 0.5$-$1.0 keV &  & \\
         & C = count rate in energy band 1.0$-$2.0 keV &  & \\
         SC\_HR3 & Hardness ratio 3 & XMM & X-ray band \\
         & Definition: hr3 =(D $-$ C)/(D $+$ C), where &  & \\
         & C = count rate in energy band 1.0$-$2.0 keV &  & \\
         & D = count rate in energy band 2.0$-$4.5 keV &  & \\
         SC\_HR4 & Hardness ratio 4 & XMM & X-ray band \\
         & Definition: hr4 = (E $-$ D)/(E $+$ D), where &  & \\
         & D = count rate in energy band 2.0$-$4.5 keV &  & \\
         & E = count rate in energy band 4.5$-$12.0 keV &  & \\
         SC\_EXTENT & Source extent & XMM & X-ray band \\
         $log(f8)$ & total flux & XMM & X-ray band \\
         $log(fx)$ & X-ray flux during 0.5-4.5 keV & XMM & X-ray band \\
         $log(f8/fr)$ & total-to-optical-flux ratio & SDSS, XMM & Optical and X-ray bands \\
         $log(fx/fr)$ & X-ray-to-optical-flux ratio & SDSS, XMM & Optical and X-ray bands \\
         psfMag\_u & $u$ PSF magnitude & SDSS & Optical band \\
         psfMag\_g & $g$ PSF magnitude & SDSS & Optical band \\
         psfMag\_r & $r$ PSF magnitude & SDSS & Optical band \\
         psfMag\_i & $i$ PSF magnitude & SDSS & Optical band \\
         psfMag\_z & $z$ PSF magnitude & SDSS & Optical band \\
         petroMag\_u & $u$ Petrosian magnitude & SDSS & Optical band \\
         petroMag\_g & $g$ Petrosian magnitude & SDSS & Optical band \\
         petroMag\_r & $r$ Petrosian magnitude & SDSS & Optical band \\
         petroMag\_i & $i$ Petrosian magnitude & SDSS & Optical band \\
         petroMag\_z & $z$ Petrosian magnitude & SDSS & Optical band \\
         $W1$ & $W1$ magnitude & CatWISE2020 & Infrared band \\
         $W2$ & $W2$ magnitude & CatWISE2020 & Infrared band \\
         \hline
    \end{tabular}%
    }
\end{table*}

%\documentclass{aastex631}
%\usepackage{graphicx}

%\begin{document}

%\end{document}

\begin{figure*}
    \centering
    
    \includegraphics[width=0.75\textwidth]{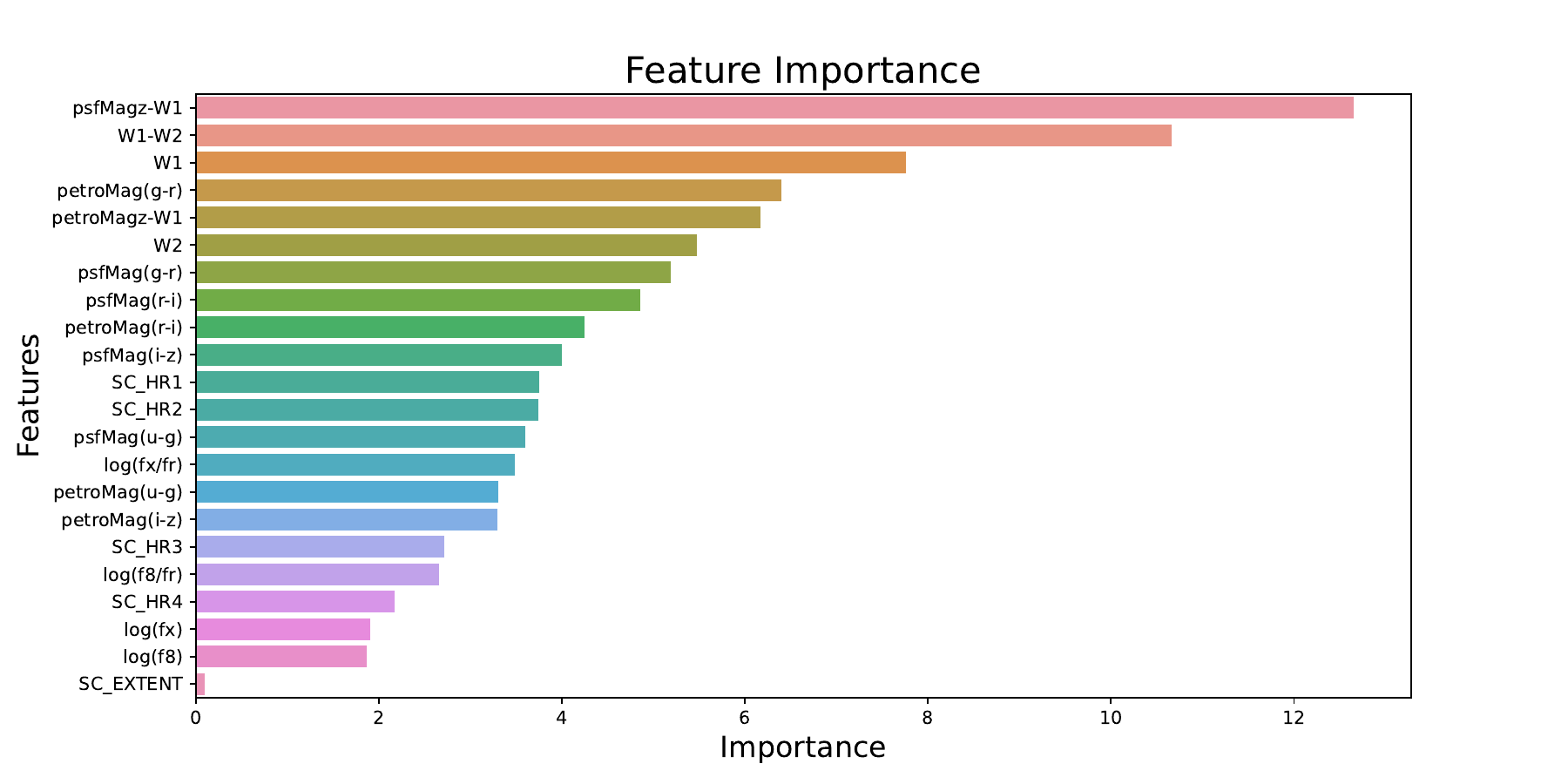}
    
    \hfill
    
    \includegraphics[width=0.75\textwidth]{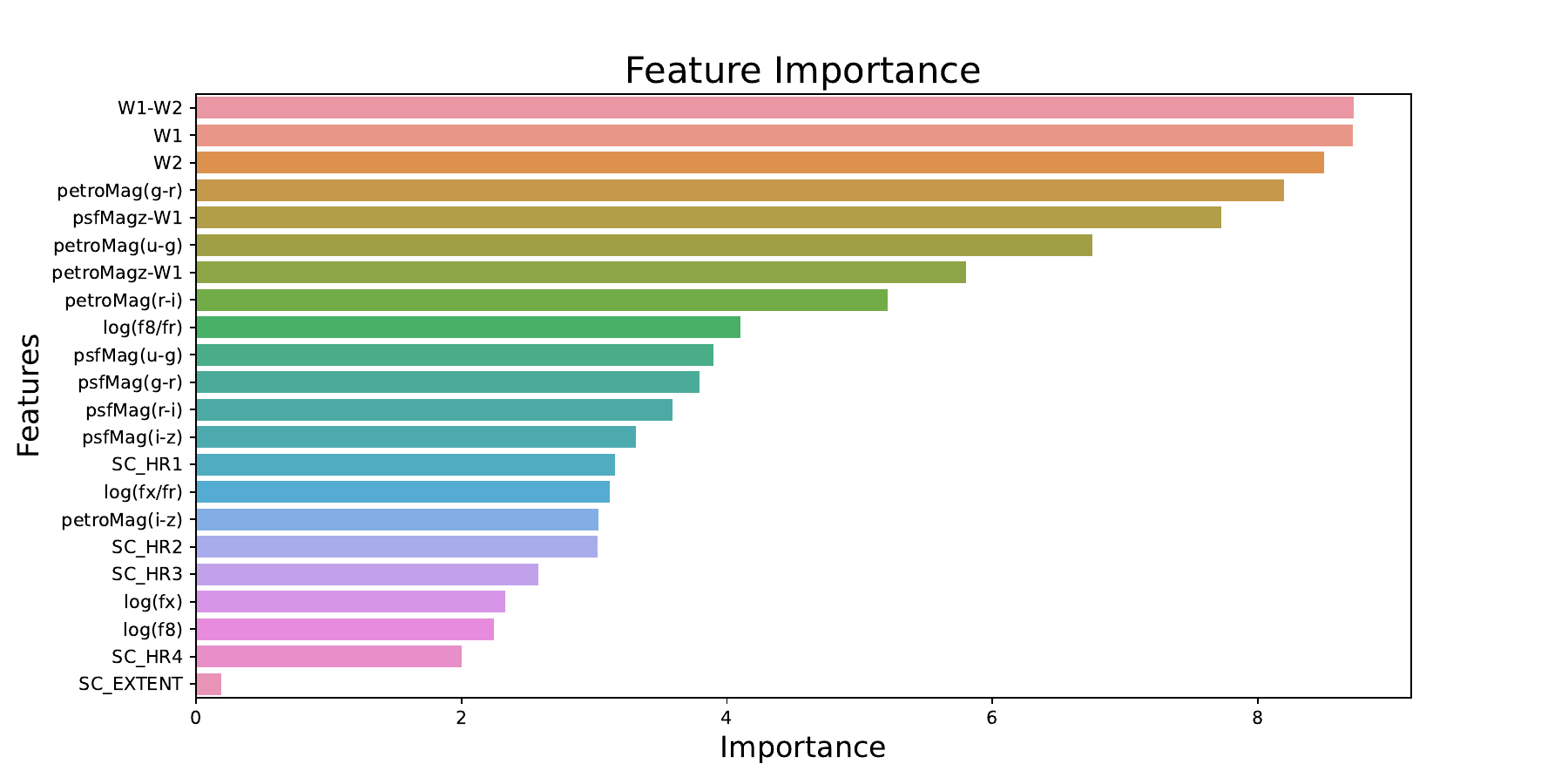}
    
    \caption{The feature importance of the XMM-WISE-SDSS sample. The above panel is feature importance by CatBoost, and the below panel is feature importance by SPE.}
    \label{Figure 2}
\end{figure*}

\section{Method}
\subsection{CatBoost}
Gradient Boosting is a powerful machine learning technique renowned for delivering state-of-the-art results across a wide array of real-world tasks. CatBoost emerges as an advanced implementation of the Gradient Boosted Decision Trees (GBDT) algorithm framework. It operates on a symmetric decision tree algorithm characterized by fewer parameters, robust support for categorical variables, and outstanding accuracy \citep{dorogush2018catboost}.
CatBoost excels in handling categorical features, gradient biases, and prediction drift issues. One of CatBoost's significant advantages lies in its innovative approach to tree structure selection, utilizing a novel method to compute leaf values, thereby mitigating overfitting risks. Specifically, CatBoost introduces the concept of ordered boosting, which is a modification of the standard gradient boosting algorithm. It avoids the problem of target leakage and reduces the prediction bias \citep{prokhorenkova2017catboost}. 

Additionally, CatBoost offers both CPU and GPU implementations, with the GPU version notably faster than counterparts like XGBoost and LightGBM. This speed advantage is a result of CatBoost's GPU support for multiple GPUs and its ability to parallelize distributed tree learning across samples or features.
Significantly, CatBoost also boasts a high-performance CPU scoring implementation, surpassing XGBoost and LightGBM in similar-sized integration. This combined superiority in handling categorical variables, optimizing gradient biases, and its efficient CPU and GPU implementations cements CatBoost as a leading choice in the realm of Gradient Boosting.

\subsection{Self-paced Ensemble}
As depicted in Table \ref{Table 1}, there is a notable imbalance in the distribution of data types, particularly evident in the relatively small number of YSOs compared to other categories. Self-paced Ensemble (SPE) is tailor-made for tackling large-scale and highly imbalanced classification tasks. By integrating undersampling and integration strategies into serial training boosting-like procedures, SPE crafts an additive model optimized for such challenging scenarios. In the initial stages of SPE training, the undersampling strategy diligently equalizes the contribution of each data bucket to the classification complexity \citep{liu2020self}, fostering an environment where even if the samples within each bucket sum up to the same value after resampling, their impact is balanced. The primary advantage of this approach lies in its adeptness at magnifying the significance of boundary samples while dampening the influence of noise on the learning trajectory. This strategic balancing act significantly enhances the model's capacity to accurately classify highly imbalanced datasets, mirroring the complexities encountered in our study's dataset. Through this approach, SPE emerges as a powerful tool for navigating the intricacies of imbalanced classification tasks, offering a  solution to real-world challenges in the scientific domain.

\subsection{Feature Selection and Model Parameter Optimization}
When employing machine learning for classification tasks, the selection of features plays a crucial role in determining the accuracy of classification outcomes. In our approach to selecting optimal features for the model, we input all features and utilize different machine learning methods, specifically CatBoost and SPE, as operational models to rank the importance of different features. Figure~\ref{Figure 2} illustrates the ranking diagram showcasing the importance of features by means of CatBoost and SPE.

An analysis of these figures reveals that both models exhibit similar preferences for the top-ranked important features. Based on this feature ranking, we initially train the model with all features and subsequently reduce one feature at a time during training. We evaluate the performance metrics such as accuracy, recall and precision as the number of features decreases. We observe slight fluctuations in performance as the number of features changes, leading us to select the best feature input based on different performance feedback criteria.

The optimization of model parameters is a very important and tedious task. For CatBoost optimization, we focus on parameters such as maximum tree depth (depth), maximum number of trees (iterations), and learning\_rate, while utilizing default values for other parameters. The depth of the tree determines the complexity of the model. The number of trees represents the number of trees to be constructed. The learning\_rate determines the contribution of each tree to the final prediction. The adjustment of these parameters is to find a balance between model complexity and generalization ability. For other parameters, such as regularization terms, feature sampling rates, etc., the default settings of CatBoost usually perform well on a variety of data. We conduct each training session with 5-fold cross-validation to derive average recall, accuracy, precision, F1-score, and runtime metrics. The optimization method proposed by \citet{li2023photometric} guides us in optimizing depth, iterations, and learning rate, resulting in optimal parameters of depth=6, iterations=3000, and learning\_rate=0.02 for our model. On the other hand, achieving good performance with SPE involves selecting Random Forest as the base classifier while maintaining default values for other model parameters. This streamlined approach allows SPE to attain a favorable performance outcome.

\subsection{Evaluation Metric}
Usually accuracy, precision, recall, F1-score are adopted as the evaluation metrics of machine learning performance. Accuracy, being an intuitive measure, is often considered a reliable evaluation index. However, it's important to note that high accuracy doesn't always equate to a good algorithm. Therefore, in the context of imbalanced astronomical datasets, it is imperative to simultaneously evaluate average accuracy, precision, recall, and F1-score. Elevated values in these metrics are indicative of superior model performance. These comprehensive assessment metrics are particularly crucial in scenarios where there is an uneven distribution of classes within the dataset, ensuring a more accurate and reliable interpretation of celestial phenomena.

Accuracy is defined as follow:
\begin{equation}
    {\rm Accuracy} = \frac{\rm TP+TN}{\rm TP+TN+FP+FN}
    \label{Eq.1}
\end{equation}
where true positive is short for TP, false negative for FN, false positive for FP, and true negative for TN.

Precision represents the proportion of rightly classified positive examples among all those classified as positive examples.
\begin{equation}
    {\rm Precision} = \frac{\rm TP}{\rm TP+FP}
    \label{Eq.2}
\end{equation}

Recall is a measure of coverage, and the proportion of rightly classified positive examples among all true positive examples.
\begin{equation}
    {\rm Recall} = \frac{\rm TP}{\rm TP+FN}
    \label{Eq.3}
\end{equation}

F1-score provides a balanced assessment of a model's performance, especially useful in handling imbalanced datasets.
\begin{equation}
    {\rm F1-score} = 2\times\frac{ {\rm Precision} \times {\rm Recall}}{\rm Precision+Recall}
    \label{Eq.4}
\end{equation}

\begin{figure*}
    \centering
    %\begin{subfigure}[t]{0.75\textwidth}
    \includegraphics[width=0.75\textwidth]{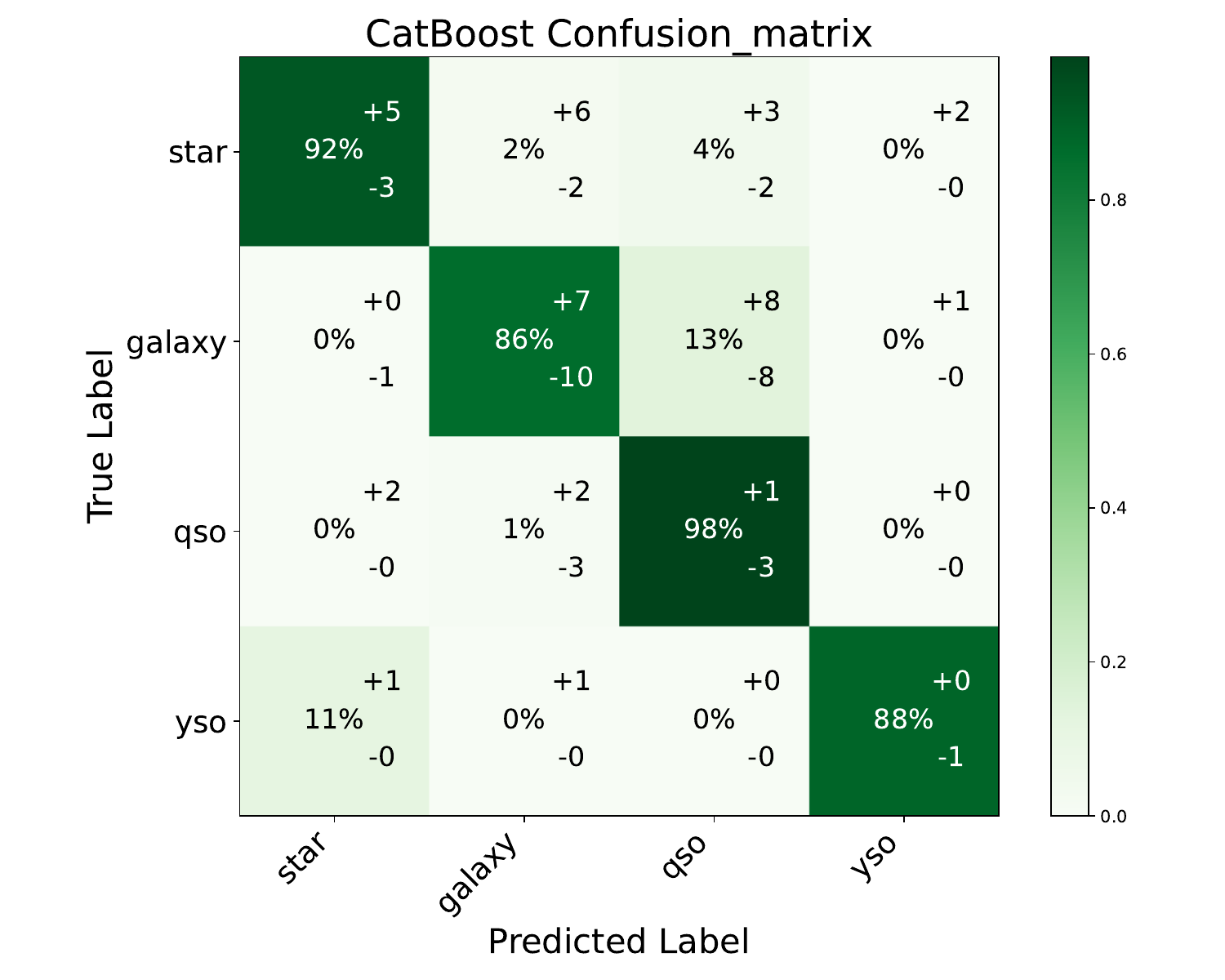}
    %\end{subfigure}
    \hfill
    %\begin{subfigure}[t]{0.75\textwidth}
    \includegraphics[width=0.75\textwidth]{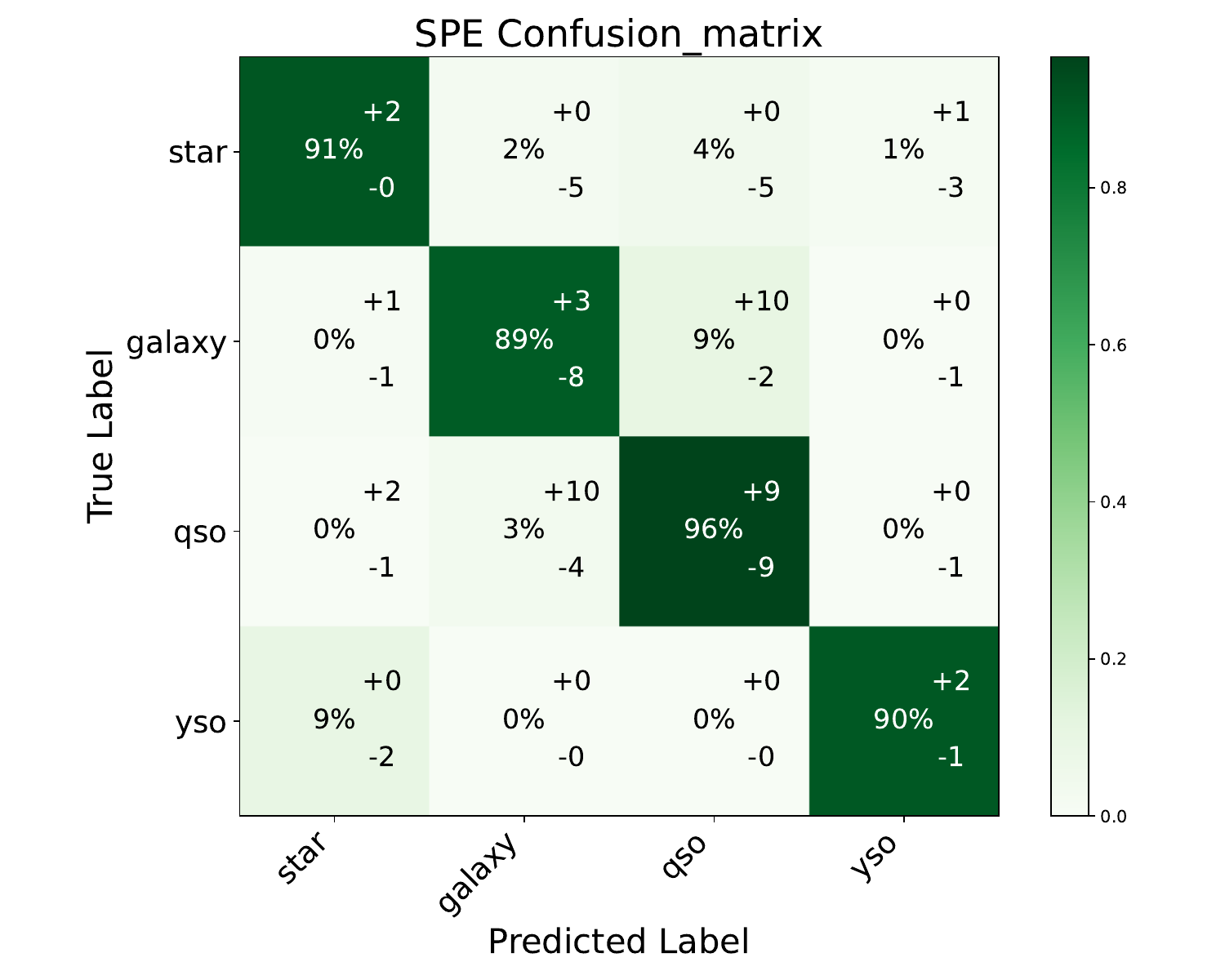}
    %\end{subfigure}
    \caption{Confusion matrix of classification for the known samples, the above panel for CatBoost while the below panel for SPE.}
    \label{Figure 3}
\end{figure*}

\section{Results and Application}
\subsection{Classification Results}
Our objective was to classify the XMM-WISE-SDSS sample and identify YSO candidates. To achieve this, we applied two machine learning algorithms: CatBoost and SPE, utilizing the model parameters obtained in Section 3.3. To assess the performance of the classifier, we employed 5-fold cross-validation. This involved dividing the samples into five parts, with four parts used for training and one part for testing, repeated five times to calculate average metric values. Confusion matrices for both machine learning algorithms are presented in Figure~\ref{Figure 3}.

The average accuracy, precision, recall, and F1-score for CatBoost are 94.437\%, 94.442\%, 94.437\%, and 94.275\%, respectively. In contrast, for SPE, they are 94.429\%, 94.555\%, 94.429\%, and 94.458\%, respectively. Analyzing these metrics alone makes it challenging to definitively determine the superiority between CatBoost and SPE. However, Figure~\ref{Figure 3} offers deeper insights. This calculation method provides a method to measure the inaccuracy of model prediction. The median confusion matrix provides a measure of the central trend, while the upper and lower bounds provide the range of possible fluctuations in the predicted value. This is very useful in evaluating the performance of the model, especially in the classification problem, which helps to understand the stability and reliability of the classification results. It illustrates that CatBoost performs exceptionally well in classifying large classes, particularly evident in its accuracy for quasars (99\% compared to SPE's 97\%). Conversely, SPE demonstrates remarkable proficiency in categorizing small classes within highly imbalanced datasets, notably excelling in the classification of YSOs (90\% compared to CatBoost's 81\%). This enhancement in the accuracy of small classes by SPE is achieved through a trade-off, whereby the accuracy of larger classes is slightly diminished.

Hence, CatBoost is adept at achieving elevated accuracy levels with predominant classes, whereas SPE excels in addressing minority classes. Astronomers frequently grapple with sample imbalances when tackling classification tasks, particularly when scouring expansive datasets for rare objects. In these contexts, SPE emerges as a prime candidate, leveraging its adeptness in navigating imbalanced datasets, as demonstrated by \cite{zhang2023catalog}. Given our primary objective of identifying YSO candidates, the SPE method proves to be more suitable for our purpose.

\subsection{Model Application}
The aforementioned results underscore the effectiveness of the SPE algorithm in identifying YSO candidates and the efficacy of CatBoost in selecting quasar candidates. We utilized both the CatBoost and SPE classifiers to classify the XMM-WISE-SDSS sample, and the classification outcomes are presented in Table \ref{Table 3}. By the SPE classifier, 20,277 star candidates, 60,454 galaxy candidates, 78~696 quasar candidates, and 1102 YSO candidates are yielded. By the CatBoost classifier, 19,233 star candidates, 49,849 galaxy candidates, 90,916 quasar candidates, and 547 YSO candidates are obtained. For researchers interested in quasars, the classification results generated by CatBoost may be preferred, whereas those focusing on YSOs may find the results produced by SPE more relevant. For the completeness of YSOs or quasars, we may adopt the combined classification result of CatBoost and SPE. Regarding accuracy, we may adopt the cross-matched classification result of CatBoost and SPE.

\begin{table*}
    \centering
    \setlength{\tabcolsep}{4mm}
    \caption{The classification result of the XMM-WISE-SDSS sample by CatBoost and SPE. Class\_CatBoost represents class given by CatBoost, Class\_SPE shows class obtained by SPE; P\_CatBoost is the classification probability by CatBoost,  P\_SPE is the classification probability by SPE. This whole table is available at http://paperdata.china-vo.org/mxy/table3.csv.} 
    \label{Table 3}
    \begin{tabular}{clllclc}
        \hline
        SRCID & SC\_RA & SC\_DEC & Class\_CatBoost & P\_CatBoost& Class\_SPE & P\_SPE\\
        \hline
        206516803010002 & 281.636194 & -2.353715 & STAR & 0.543& YSO & 0.784\\
        201645602010018 & 86.785682 & 0.00457154 & STAR & 0.617& YSO & 0.632\\
        207956701010042 & 226.654933 & 3.638037 & QSO & 0.904& QSO & 0.536\\
        200114201010045 & 100.278052 & 9.7909098 & YSO & 0.638& YSO & 0.765\\
        206727201010003 & 122.242080 & 20.812036 & STAR & 0.998& STAR & 0.941\\
        202033905010028 & 166.078129 & 40.7661459 & GALAXY & 0.967& GALAXY & 0.543\\
        202033906010006 & 229.405224 & 42.624143 & QSO & 0.989& QSO & 0.710\\
        200114202010078 & 100.4326076 & 9.6805009 & YSO & 0.694& YSO & 0.812\\
        200114201010236 & 100.356461 & 9.734582 & YSO & 0.599& YSO & 0.775\\
        200019303010005 & 193.639854 & 10.337648 & QSO & 0.979& QSO & 0.673\\
        200001101010011 & 64.905070 & 56.064407 & GALAXY & 0.573& GALAXY & 0.369\\
        206527201010054 & 216.101248 & 26.452912 & GALAXY & 0.892& GALAXY & 0.618\\
        201002410010027 & 357.389444 & 36.657394 & QSO & 0.996& QSO & 0.586\\
        206779807010043 & 202.895200 & 47.196223 & QSO & 0.735& GALAXY & 0.483\\
        203057506010036 & 172.612143 & 0.772320 & QSO & 0.940& QSO & 0.618\\
        208218713010048 & 200.874507 & 31.448973 & GALAXY & 0.874& GALAXY & 0.531\\
        205035601010043 & 84.971938 & -7.3324481 & STAR & 0.649& YSO & 0.737\\
        201118704010007 & 239.612559 & 27.216867 & STAR & 0.989& STAR & 0.626\\
        208611704010009 & 343.375679 & 62.306190 & STAR & 0.580& YSO & 0.663\\
        \hline
    \end{tabular}
\end{table*}

\begin{figure*}
    \centering
    \includegraphics{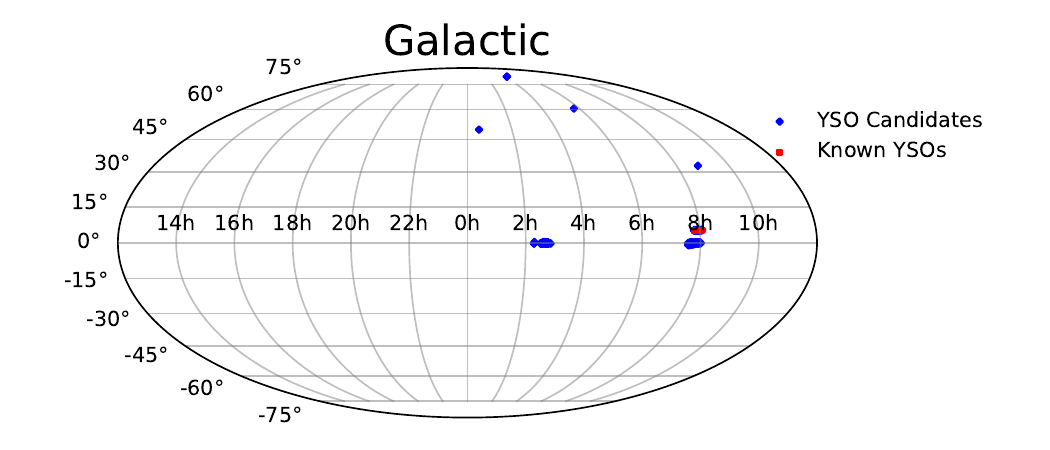}
    \caption{The known YSOs and YSO candidates in Galactic coordinates. The blue diamonds represent YSO candidates selected by SPE, and the red squares represent known YSOs.}
    \label{Figure 4}
\end{figure*}

\begin{figure*}
    \centering
    \includegraphics[width=0.9\linewidth]{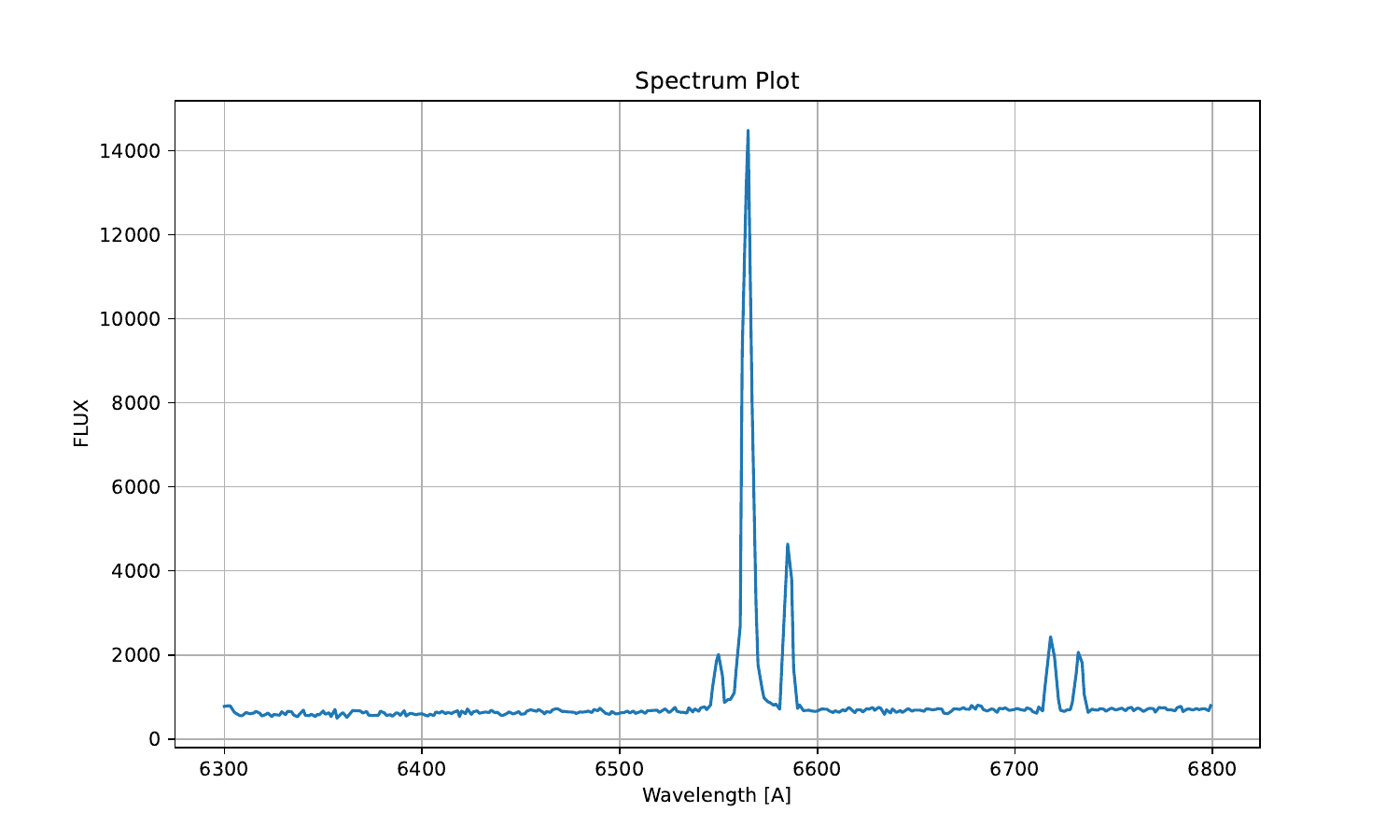}
    \caption{The spectrum of LAMOST LRS spec-56617-VB081S05V2\_sp06-073.fits. It shows strong H$ \alpha $ emission at 6562.8 Å.}
    \label{Figure 5}
\end{figure*}

\begin{figure*}
    \centering
    \includegraphics[width=0.9\linewidth]{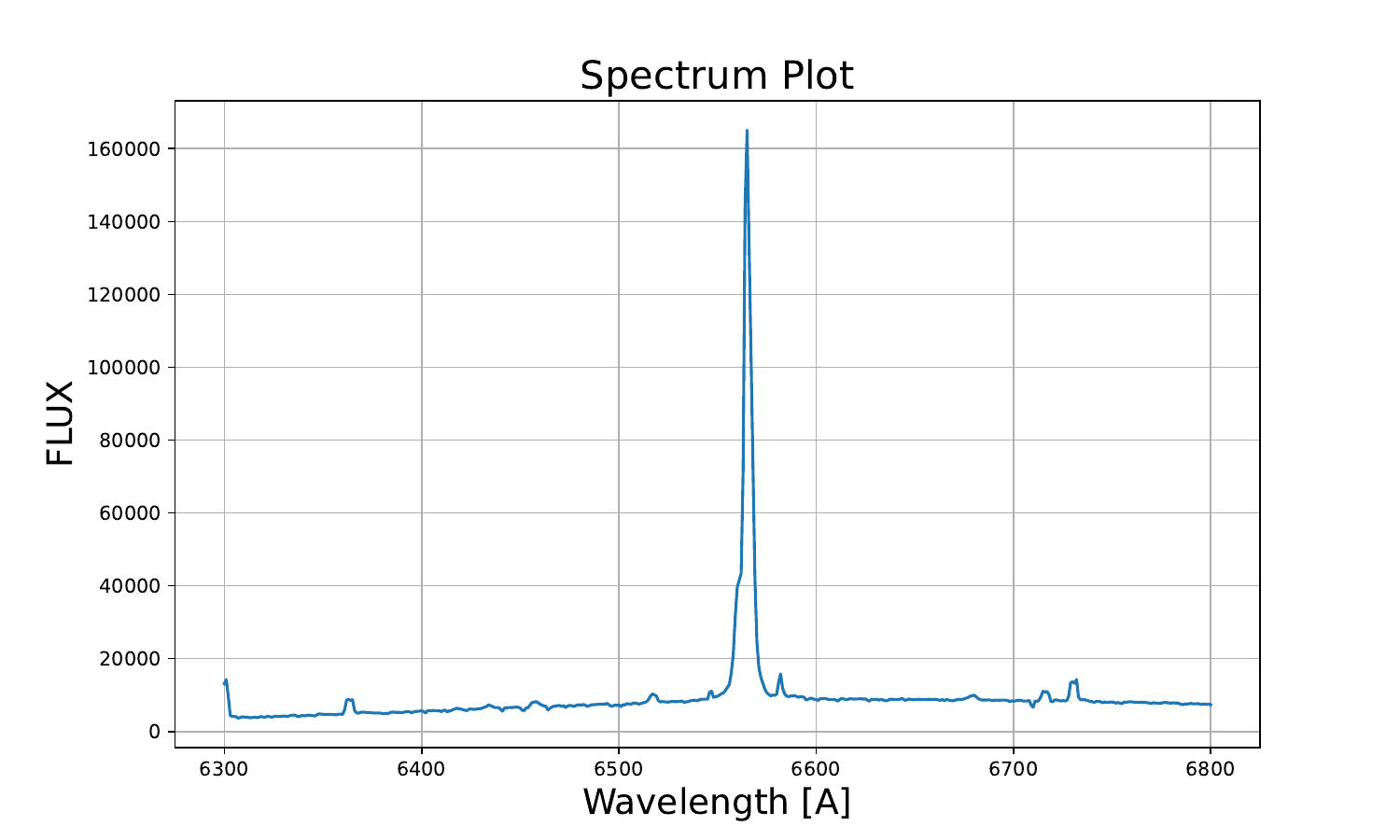}
    \caption{The spectrum of LAMOST MRS med-59980-TD043724N254338B01\_sp14-231.fits. It shows strong H$ \alpha $ emission at 6562.8 Å.}
    \label{Figure 6}
\end{figure*}

SPE algorithms identified 1102 potential YSO candidates in the XMM-WISE-SDSS dataset, of which 258 were known YSOs. To further validate these YSO candidates, we analyzed their spatial distribution by plotting the locations of both the candidates and known YSOs in the sky, as illustrated in Figure~\ref{Figure 4}. It is evident that the majority of YSOs are concentrated along the Galactic plane, with only a few located at higher latitudes. Additionally, we cross-matched the YSO candidates with the LAMOST DR11 catalog within a radius of 3 arc seconds, yielding 179 sources that exhibited high-quality spectra with distinct YSO spectral features. We also performed a cross-match of YSO candidates within a 3 arc second radius in the SIMBAD database, resulting in the identification of 268 candidates. Among these, 22 YSOs were consistently identified in both the LAMOST and SIMBAD databases. Additionally, we conducted a cross-match of YSO candidates within a 3 arc second radius in the VizieR database, identifying 7 additional YSO candidates. Overall, more than half of the 1,102 YSO candidates were confirmed as YSOs. Consequently, the remaining 412 YSO candidates remain unidentified and require further observational efforts for follow-up confirmation.
%We also performed a cross-match of the YSO candidates on the SIMBAD database within a 3 arcsecond radius, identifying 287 candidates as YSOs. Of these, 37 YSOs were commonly identified in both the LAMOST and SIMBAD databases. Among these 1102 YSO candidates, over half have been confirmed as YSOs. Consequently, the remaining 415 YSO candidates are unidentified, necessitating further observational follow-up for confirmation.
Figures.~\ref{Figure 5} and \ref{Figure 6} display spectra of two such YSOs from the LAMOST database, exemplifying the 179 sources. As YSOs, these stars exhibit frequent surface and atmospheric activity, including stellar flares. These activities can enhance the ionization of hydrogen atoms and increase the frequency and intensity of electron transitions between energy levels, which in turn results in more pronounced H$\alpha$ emission lines.
These spectral signatures are prominently visible in Figures.~\ref{Figure 5} and \ref{Figure 6}, taking the two medium-resolution spectra of LAMOST for example.

\section{CONCLUSIONS}
Based on the distribution of stars, galaxies, quasars, and YSOs within two-dimensional characteristic spaces, distinguishing them poses considerable challenges, especially between stars and YSOs. However, upon examining the 2D figures, it becomes apparent that these features exert varying degrees of influence on classification. Thus, by leveraging X-ray signatures, infrared, and optical information, we employed two classifiers, CatBoost and SPE, to effectively classify stars, galaxies, quasars, and YSOs while addressing unbalanced classification issues. 

These findings highlight the superiority of the SPE algorithm over CatBoost, especially in scenarios with highly imbalanced classification features. The SPE algorithm shows excellent ability in classifying minority classes, and the stability and reliability of the model are better than CatBoost when classifying minority classes, although this is at the expense of the accuracy of large classes. It is particularly noteworthy that it shows superior performance in classifying the studied YSOs.

Future follow-up studies can focus on identifying these previously unidentified YSO candidates, thereby enriching the YSO sample repository. A more extensive YSO sample database is invaluable for comprehensively studying star formation regions and their underlying processes. Moreover, the SPE algorithm's exceptional performance in identifying YSOs positions it as a promising tool for exploring additional databases to uncover more YSO candidates in newer datasets. In addition, SPE may be applied to other imbalance classification problems in astronomy. 

\section*{Acknowledgements}
We are very grateful to the referee's comments and suggestions. This paper is funded by the National Natural Science Foundation of China under grants No.12273076, No.12203077, No.12133001 and No.12373110. This research has made use of data obtained from the 4XMM XMM–Newton serendipitous source catalogue compiled by the 10 institutes of the XMM–Newton Survey Science Centre selected by ESA. The computing task of this paper was carried out on the computing platform at China National Astronomical Data Center (NADC). NADC is a National Science and Technology Innovation Base hosted at National Astronomical Observatories, Chinese Academy of Sciences. This publication makes use of data products from the Wide-field Infrared
Survey Explorer, which is a joint project of the University of California, Los Angeles, and the Jet Propulsion Laboratory/California Institute of Technology, funded by the National Aeronautics and Space Administration. The Guoshoujing Telescope (the Large Sky Area Multi-object Fiber Spectroscopic Telescope, LAMOST) is a National Major Scientific Project built by the Chinese Academy of Sciences. Funding for the project has been provided by the National Development and Reform Commission. LAMOST is operated and managed by the National Astronomical Observatories, Chinese Academy of Sciences. This research has made use of the SIMBAD database,
operated at CDS, Strasbourg, France. This research also uses the VizieR database.

We acknowledgment SDSS data bases. Funding for the Sloan Digital Sky Survey IV has been provided by the Alfred P. Sloan Foundation, the US Department of Energy Office of Science, and the Participating Institutions. SDSS-V acknowledges support and resources from the Center for High-Performance Computing at the University of Utah. The SDSS web site is \href{http://www.sdss.org}{www.sdss.org}. SDSS-V is managed by the Astrophysical Research Consortium for the Participating Institutions of the SDSS Collaboration including the Brazilian Participation Group, the Carnegie Institution for Science, Carnegie Mellon University, the Chilean Participation Group, the French Participation Group, the Harvard-Smithsonian Center for Astrophysics, Instituto de Astrof´ısica de Canarias, the Johns Hopkins University, Kavli Institute for the Physics and Mathematics of the Universe (IPMU)/University of Tokyo, Lawrence Berkeley National Laboratory, Leibniz Institut f¨ur Astrophysik Potsdam (AIP), Max-Planck-Institut f¨ur Astronomie (MPIA Heidelberg), Max-Planck-Institut f¨ur Astrophysik (MPA Garching), Max-Planck-Institut f¨ur Extraterrestrische Physik (MPE), National Astronomical Observatories of China, New Mexico State University, New York University, University of Notre Dame, Observat´ario Nacional/MCTI,the Ohio State University, Pennsylvania State University, Shanghai Astronomical Observatory, the United Kingdom Participation Group, Universidad Nacional Aut´onoma de M´exico, University of Arizona,University of Colorado Boulder, University of Oxford, University of Portsmouth, University of Utah, University of Virginia, University of Washington, University of Wisconsin, Vanderbilt University, and Yale University.

\section*{Data Availability}
The predicted 4XMM-DR13 catalogue is available on paperdata at http://paperdata.china-vo.org, and can be accessed at http://paperdata.china-vo.org/mxy/table3.csv. The part of it is shown in Table~3.

\bibliography{sample631}{}
\bibliographystyle{aasjournal}



\end{document}